\documentclass[journal=jctcce,manuscript=article]{achemso}
\setkeys{acs}{maxauthors=0,articletitle=true}

\usepackage{achemso}
\usepackage{placeins}
\usepackage{graphics}
\usepackage{amssymb,amsfonts}
\usepackage{mathtools}
\usepackage{physics}
\usepackage{subcaption}
\usepackage{graphicx}
\usepackage[table,dvipsnames]{xcolor}
\usepackage{multirow}
\usepackage{caption}
\usepackage{booktabs}
\usepackage{colortbl}
\usepackage{amsmath}
\usepackage{amsopn}
\usepackage{bm}
\usepackage{color}
\usepackage{array}
\usepackage{lscape}
\usepackage{mciteplus}
\usepackage[version=3]{mhchem}
\usepackage[normalem]{ulem}
\usepackage{listings}
\usepackage{enumerate}
\usepackage{lmodern}

\author{Jonas Greiner}
\affiliation[mainz]{Department Chemie, Johannes Gutenberg-Universit{\"a}t Mainz\\Duesbergweg 10--14, 55128 Mainz, Germany}
\altaffiliation{These authors contributed equally}
\author{Ivan Gianni}
\altaffiliation{These authors contributed equally}
\author{Tommaso Nottoli}
\author{Filippo Lipparini}
\email{filippo.lipparini@unipi.it}
\affiliation{Dipartimento di Chimica e Chimica Industriale, Universit\`a di Pisa\\Via G. Moruzzi 13, Pisa, 56124, Italy}
\author{Janus J. Eriksen}
\email{janus@dtu.dk}
\affiliation[dtu]{DTU Chemistry, Technical University of Denmark\\Kemitorvet Bldg. 206, 2800 Kgs. Lyngby, Denmark}
\author{J{\"u}rgen Gauss}
\email{gauss@uni-mainz.de}
\affiliation[mainz]{Department Chemie, Johannes Gutenberg-Universit{\"a}t Mainz\\Duesbergweg 10--14, 55128 Mainz, Germany}

\title{MBE-CASSCF Approach for the Accurate Treatment of Large Active Spaces}

\begin{document}

\begin{tocentry}
    \includegraphics[width=\linewidth]{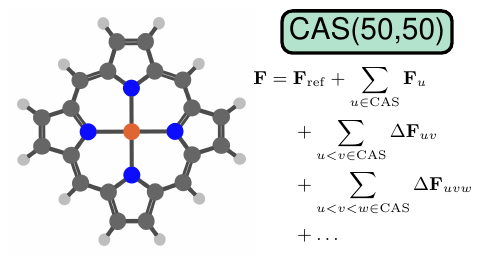}
\end{tocentry}

\begin{abstract}

We present a novel implementation of the complete active space self-consistent field (CASSCF) method that makes use of the many-body expanded full configuration interaction (MBE-FCI) method to incrementally approximate electronic structures within large active spaces. On the basis of a hybrid first-order algorithm employing both Super-CI and quasi-Newton strategies for the optimization of molecular orbitals, we demonstrate both computational efficacy and high accuracy of the resulting MBE-CASSCF method. We assess the performance of our implementation on a set of established numerical tests before applying MBE-CASSCF in the investigation of the triplet-quintet spin gap of iron(II) porphyrin with active spaces as large as 50 electrons in 50 orbitals.

\end{abstract}

\section{Introduction}

Commonly used approximations to the solution of the electronic Schrödinger equation in quantum chemistry are typically based on the assumption that the electronic ground-state wave function can be satisfactorily described using a single Slater determinant. This premise is often perfectly reasonable as long as said state is energetically well-separated from any other electronic states in the molecular system, but many systems of broad chemical interest, such as, conjugated hydrocarbons and transition-metal complexes, may require the use of multiple Slater determinants to yield a qualitatively correct description of the involved states. In addition, both bond-breaking processes and electronic transitions often result in 
electronic states that exhibit contributions from nearly equivalent determinants, rendering the choice of a single reference somewhat arbitrary.\cite{mest} The \textit{de facto} standard today for the computational study of such phenomena is the complete active space self-consistent field (CASSCF) method,\cite{Roos1980, roos1980_sci, siegbahn1980_comp, siegbahn1981_sci, Roos2007} which optimizes both a linear combination of Slater determinants and the molecular orbitals (MOs) used to construct them. The set of determinants is defined by dividing the complete set of MOs into a doubly occupied inactive set, a virtual set of unoccupied MOs, and a set of active MOs, in the space of which a given number of active electrons are distributed.\\

Applying the CASSCF method to a chemical problem is challenging for two main reasons. First, results will depend significantly on the choice of the active space,\cite{Roos2007} typically necessitating some {\textit{a priori}} chemical knowledge of the electronic structure of the molecular system in question and thus introducing a degree of arbitrariness. To mitigate this problem somewhat, automated strategies for choosing active spaces have been proposed over the years.\cite{Bofill1989, Veryazov2011, Keller2015, Stein2016, Sayfutyarova2017, Sayfutyarova2019} Second, and perhaps foremost, the computational cost of CASSCF calculations scales exponentially with the size of the active space, and optimal choices---yielded by either human intuition or an automatic procedure---may hence easily become practically intractable.\\

A CASSCF wave function is determined by optimizing the electronic energy with respect to both MO and configuration interaction (CI) coefficients. In first-order procedures, the optimizations of these two parameter sets are decoupled in an alternating gradient fashion. In other words, a complete active space configuration interaction (CASCI) solution is computed alongside its resulting one- and two-body reduced density matrices (RDMs), which are then used to perform an optimization step on the orbitals in an iterative procedure.
For increasingly larger active spaces, the CASCI step soon becomes the limiting factor in CASSCF, due to its combinatorial scaling with respect to the number of active MOs. In practice, active spaces beyond 18 electrons distributed within 18 orbitals are computationally intractable on standard hardware, as the configurational space grows into the billions of determinants, with active spaces as large as 22 electrons in 22 orbitals being treatable only on large-scale parallel computers.\cite{Vogiatzis2017} Initial attempts at mitigating this problematic scaling were successful by restricting the number of relevant configurations within the active space.\cite{malmqvist1990,Ma2011,Vogiatzis2017} Alternatively, approximate CI methods have emerged as a promising tool to tame the scaling of this step in CASSCF, and these may, for the sake of brevity, be conveniently classified into four main categories:\cite{Eriksen2020b} adaptive sampling and general selected CI (SCI) methods~\cite{Smith2017,levine2020casscf,Guo2021}, methods based on tensor-decomposition techniques, namely, density-matrix renormalization group theory~\cite{Zgid2008,Ghosh2008,Yanai2010,Kurashige2011,Nakatani2017,Ma2017}, stochastic quantum Monte-Carlo methods~\cite{Manni2016,Dobrautz2021}, and, most important in the context of the present work, incremental methods based on the many-body expansion (MBE), as in the works on the iCASSCF method by Zimmerman and co-workers~\cite{Zimmerman2019,Dang2021}. Other noteworthy approaches for treating large active spaces are the variationally orbital-optimized CC approaches,\cite{Krylov1998,Parkhill2010,Parkhill2010a,Lehtola2017} the active space decomposition method,\cite{Parker2013,Parker2014,Kim2015} the cluster mean-field method,\cite{JimenezHoyos2015} and the localized active space SCF method.\cite{Hermes2019,Hermes2020,Pandharkar2022}\\

In a similar vein, the present work will be concerned with the coupling of an optimized implementation of standard CASSCF to the recent many-body expanded full configuration interaction (MBE-FCI) method.\cite{Eriksen2017,Eriksen2018,Eriksen2019,Eriksen2019a,Eriksen2020, Eriksen2021} In MBE-FCI, conventional FCI is incrementally approximated by performing a multitude of CASCI calculations within truncated orbital spaces, that is, without recourse to an explicit sampling of the full wave function. More specifically, properties such as (excitation) energies and (transition) dipole moments of arbitrary states~\cite{Eriksen2020} are all decomposed into contributions from sets of optimized, spatially localized orbitals, subject to an efficient screening protocol that, in turn, leads to a reduced number of increment calculations throughout the MBE. A recent benchmark assessment of leading approximations to FCI convincingly demonstrated both the accuracy, reliability, and applicability of the MBE-FCI approach.\cite{Eriksen2020a} In the present work, we outline our implementation of a hybrid method for CASSCF orbital optimizations involving an approximate, embarrassingly parallel CASCI solver based on MBE-FCI. The use of MBE-FCI within a CASSCF implementation requires MBEs of either RDMs or generalized Fock matrices, both tensor objects rather than simple scalar quantities. Our algorithm differs from other incremental implementations in that it is not limited to molecular valence regions, but may be applied to arbitrary active spaces. Additionally, we propose the use of a closed-form, recursive implementation of the MBE that enables memory-efficient orbital optimizations, while evaluations of final energies can be refined through the use of orbital-based screening. To demonstrate the applicability of the resulting algorithm, we apply MBE-CASSCF to investigate the energy gap between the two lowest-lying electronic states of triplet and quintet spin of iron(II) porphyrin within active spaces as large as 50 electrons in 50 orbitals.

\section{Theory}

We start by briefly reviewing the main quantities required in the orbital optimization step of CASSCF with particular emphasis on details relating to a coupling to the MBE-FCI method.

\subsection{CASSCF Orbital Gradient}

In treating large active spaces with approximate FCI solvers, it is of paramount importance to avoid explicitly building and storing the CASCI wave function in its entirety. As such, a decoupling of the CI and MO optimization steps is essential, allowing for the communication between the two degrees of freedom to be handled via RDMs. Important in this context is the fact that RDMs make it possible to compute the gradient of the energy with respect to orbital rotations, as will be detailed in the following. Throughout, indices $\{i, j, k, \dots\}$, $\{u, v, w, \dots\}$, and $\{a, b, c, \dots\}$ will be used to label inactive, active, and external orbitals, respectively, with respect to the CASSCF active space. As is convention, the wave function,
\begin{equation} \label{wf_ansatz}
    \ket{\Psi} = e^{-\hat{\kappa}}\ket{0} \ ,
\end{equation}
is parametrized with respect to orbital rotations using an exponential operator, $\exp(-\hat{\kappa})$. The antisymmetric operator,
\begin{equation}
    \hat{\kappa} = \sum_{p<q}\kappa_{pq}\left(\hat{E}_{pq} - \hat{E}_{qp}\right) \ , \label{kappa_eq}
\end{equation}
ensures that the transformation is unitary while preserving orthogonality of the orbitals. In Eq. \ref{kappa_eq}, $\hat{E}_{pq}$ is a singlet excitation operator and $\ket{0}$ is the multiconfigurational reference wave function. Following this Ansatz, the CASSCF energy,
\begin{equation} \label{casscf_energy}
    E = \sum_{uv}D_{uv}F^I_{uv} + \sum_{uvxy}d_{uvxy}(uv|xy) + 2\sum_i\left(h_{ii} + F^I_{ii}\right) + E_\mathrm{nuc} \ ,
\end{equation}
can be written in terms of one- ($D_{uv}$) and two-body ($d_{uvxy}$) RDMs restricted to the active orbitals, the inactive Fock matrix,
\begin{equation} \label{inact_fock_matrix}
    F^I_{pq} = h_{pq} + \sum_i\left[2(pq|ii) - (pi|iq)\right] \ ,
\end{equation}
two-electron integrals, $(pq|rs)$, and the nuclear repulsion energy, $E_\mathrm{nuc}$. The inactive Fock matrix in Eq. \ref{inact_fock_matrix} additionally requires the standard one-electron integrals, $h_{pq}$. The orbital-rotation coefficients, $\kappa_{pq}$, can be obtained by varying the orbitals such that the orbital gradient vanishes with a dedicated first-order solver, such as, Super-CI~\cite{roos1980_sci, siegbahn1981_sci,malmqvist1990,kollmar2019_ptsci,angeli2002_ptsci} or Quasi-Newton.\cite{fischer1992_qn,chaban1997_qn} Both require the determination of the orbital gradient itself. Since the only nonredundant rotations in a conventional CASSCF calculation are those among different classes of orbitals, the nonvanishing blocks of the orbital gradient are:
\begin{subequations}
\label{gradient_eqs}
\begin{align}
    & g_{ia} = 2F_{ia}\label{gradient1}\\
    & g_{iu} = 2\left(F_{iu} - F_{ui}\right)\label{gradient2}\\
    & g_{ua} = 2F_{ua} \ . \label{gradient3}
\end{align}
\end{subequations}
In Eqs. \ref{gradient_eqs}, $F_{pq}$ are elements of the generalized Fock matrix
\begin{subequations}
\label{gf_eqs}
\begin{align}
    & F_{ip} = 2\left(F^I_{ip} + F^A_{ip}\right) \label{gf1}\\
    & F_{up} = \sum_v D_{uv}F^I_{vp} + Q_{up} \label{gf2}\\
    & F_{ap} = 0 \label{gf3}
\end{align}
\end{subequations}
where
\begin{equation} \label{act_fock_matrix}
    F^A_{pq} = \sum_{uv}D_{uv}\left[(pq|uv) - \frac{1}{2}(pv|uq)\right]
\end{equation}
is the corresponding active Fock matrix, and the auxiliary matrix in Eq. \ref{gf2} reads as follows
\begin{equation} \label{aux_matrix}
    Q_{up} = \sum_{vxy}d_{uvxy}(pv|xy) \ .
\end{equation}
In the context of the present work, the 1- and 2-RDMs as well as the generalized Fock matrix are not formally exact quantities obtained from a strict solution of the CASCI problem, but rather targets of an incomplete (truncated) MBE inside the active space. The configurational step of CASSCF is hence not completely redundant with respect to rotations amongst active orbitals, and these cannot be excluded from the orbital optimization if all parameters are to be variationally optimized. The orbital gradient thus has an additional non-vanishing term,
\begin{align}
    & g_{uv} = 2\left(F_{uv} - F_{vu}\right) \ .
\end{align}
In general, in the case of near-exact expansions, the active-active rotations in question still remain nearly redundant. In fact, their explicit optimization may lead to severe convergence issues, as will be discussed later on in Sect. \ref{act_act_rot_section}.

\subsection{Many-Body Expanded Full Configuration Interaction}

In the treatment of many-body interactions for systems where exact theory is practically infeasible, the correlation energy may be conveniently recast as an MBE in a given MO basis:\cite{Nesbet1967, Nesbet1967a, Nesbet1968}
\begin{align}\label{mbe}
    E&=E^{(1)}+E^{(2)}+E^{(3)}+E^{(4)}+\hdots \nonumber \\
    &=\sum_p\varepsilon_p+\sum_{p<q}\Delta\varepsilon_{pq}+\sum_{p<q<r}\Delta\varepsilon_{pqr}+\sum_{p<q<r<s}\Delta\varepsilon_{pqrs}+\hdots
\end{align}
Truncations of the expansion in Eq. \ref{mbe} upon convergence will prevent the number and size of individual calculations from growing beyond the limit of technical capabilities while still yielding accurate approximations. The individual increments are defined in a recursive manner on the basis of contributions from earlier orders in the expansion, and the involved energies are determined from individual CASCI calculations.\\

In principle, any molecular property can be decomposed into individual orbital contributions and thus be approximated through a truncated MBE. The crux of the matter is to find a suitable target for which the expansion converges in a sufficiently rapid fashion, which may be achieved by exploring multiple characteristics: an appropriate reference wave function, a compact (tailored) MO basis, a convenient definition of the expansion itself, etc. In MBE-FCI theory,\cite{Eriksen2017,Eriksen2018,Eriksen2019,Eriksen2019a,Eriksen2020, Eriksen2021} the complete set of MOs is divided into so-called reference and expansion spaces. The former of these two contains the set of MOs that is always included in every single increment, while the latter represents the set of MOs that ultimately define the objects of the MBE. For an optimal choice of reference space, the electron correlation between the orbitals spanning the reference space should amount to a substantial proportion of the total property under consideration. Such a choice can be motivated on the basis of either chemical intuition or prior knowledge of major contributions to the correlated wave function.

\section{Implementation}\label{impl_sect}

\subsection{MBE-FCI as CASSCF Kernel}\label{mbe_fci_casscf}

When considering MBE-FCI as a replacement for the standard CASCI solver in CASSCF, rapid MBE convergence is crucial, not merely for the accuracy of the desired property, but also for the convergence of the CASSCF optimization itself. Past investigations of MBE-FCI treatments of correlation energies have convincingly demonstrated how spatially localized orbitals accelerate convergence throughout an MBE, even for compact systems.\cite{Eriksen2019} For this reason, and due to the inherent variance on active-active MO rotations, we opt for localizing MOs within the active space based on a standard Pipek-Mezey localization procedure prior to initiating the CASSCF optimization.\cite{Pipek1989} Due to the fact that a single determinant occupied according to the Aufbau principle is generally used as the reference wave function in MBE-FCI, doubly occupied, singly occupied, and virtual subspaces are localized separately.\\

The increments needed to construct the MBE contribution at a given order are obtained by performing CASCI calculations involving the increment orbitals and subtracting increments at earlier orders such that only the mutual contribution of the increment orbitals remains. There are two options when seeking to approximate the CASSCF orbital gradient in an MBE-based framework: targeting the 1- and 2-body RDMs or targeting the generalized Fock matrix (cf. Eqs. \ref{gf_eqs}). The definitions of increments for these quantities will necessitate that the active space is further divided into occupied, active, and virtual spaces with respect to the increment active space, for which we will introduce the notation $\{i', j', k', \ldots\}$, $\{u', v', w', \ldots\}$, and $\{a', b', c', \ldots\}$, respectively. On par with standard MBE-FCI, the 1- and 2-RDMs,
\begin{align}
D_{uv}&=D_{uv}^\mathrm{ref}+D_{uv}^\mathrm{corr} \label{ref_1_rdm} \\    d_{uvwx}&=d_{uvwx}^\mathrm{ref}+d_{uvwx}^\mathrm{corr} \ , \label{ref_2_rdm}
\end{align}
may be calculated by summing reference (ref) and correlation (corr) contributions, the latter of which are expanded in terms of recursive increments:
\begin{align}
    D_{uv}^\mathrm{corr}&=\sum_{u'\in\{uv\}}\left(\mathbf{D}_{u'}^\mathrm{corr}\right)_{uv}+\sum_{u'<v'\in\{uv\}}\left(\Delta\mathbf{D}_{u'v'}^\mathrm{corr}\right)_{uv}+\sum_{u'<v'<w'\in\{uv\}}\left(\Delta\mathbf{D}_{u'v'w'}^\mathrm{corr}\right)_{uv}+\hdots \label{corr_1_rdm} \\
    d_{uvwx}^\mathrm{corr}&=\sum_{u'\in\{uvwx\}}\left(\mathbf{d}_{u'}^\mathrm{corr}\right)_{uvwx}+\sum_{u'<v'\in\{uvwx\}}\left(\Delta\mathbf{d}_{u'v'}^\mathrm{corr}\right)_{uvwx}+\sum_{u'<v'<w'\in\{uvwx\}}\left(\Delta\mathbf{d}_{u'v'w'}^\mathrm{corr}\right)_{uvwx}+\hdots \label{corr_2_rdm}
\end{align}
$\left(\mathbf{D}_{u'}^\mathrm{corr}\right)_{uv}$ describes the correlation contribution to the 1-RDM element of orbitals $u$ and $v$ obtained from the CASCI calculation involving orbital $u'$. The same is true for $\left(\Delta\mathbf{D}_{u'v'}^\mathrm{corr}\right)_{uv}$, except that the correlation contribution is now obtained from a CASCI calculation involving orbitals $u'$ and $v'$, and that the previous-order increments $\left(\mathbf{D}_{u'}^\mathrm{corr}\right)_{uv}$ and $\left(\mathbf{D}_{v'}^\mathrm{corr}\right)_{uv}$ are subtracted such that the resulting increment only describes the mutual contribution of both orbitals. For higher-order increments, all previous-order increments are subtracted. In Eqs. \ref{ref_1_rdm} and \ref{ref_2_rdm}, the reference 1- and 2-RDMs usually describe a single-determinant wave function occupied according to the Aufbau principle, in which case all elements of Eqs. \ref{corr_1_rdm} and \ref{corr_2_rdm} needed to describe MOs outside the increment active space will vanish. From a computational perspective, this will significantly reduce memory requirements in the early orders of the expansion. However, from a theoretical point of view, whenever the expansion space includes more than one occupied orbital and the MBE is terminated early on, both the energy and the orbital gradient constructed from the 1- and 2-RDMs in Eqs. \ref{corr_1_rdm} and \ref{corr_2_rdm} will depart slightly from direct expansions of these properties. The interested reader is referred to the supporting information (SI) for a more detailed discussion of this issue.\\

Alternatively, RDMs can be immediately contracted with MO integrals for every individual increment calculation to form the corresponding elements of the generalized Fock matrix in Eqs. \ref{gradient_eqs}. Due to the recursive nature of the MBE, such an expansion will require storage of the general-occupied and general-active blocks of the matrix for every increment:
\begin{align}
F_{ip}&=F_{ip}^\mathrm{ref}+F_{ip}^\mathrm{corr} \label{f_ip} \\
F_{up}&=F_{up}^\mathrm{ref}+F_{up}^\mathrm{corr} \ . \label{f_up}
\end{align}
The correlated parts of the generalized Fock matrix in Eqs. \ref{f_ip} and \ref{f_up} are defined as follows
\begin{align}
    F_{ip}^\mathrm{corr}&=&&\sum_{u'\in\mathrm{CAS}}\left(\mathbf{F}_{u'}^\mathrm{corr}\right)_{ip}+\sum_{u'<v'\in\mathrm{CAS}}\left(\Delta\mathbf{F}_{u'v'}^\mathrm{corr}\right)_{ip}+\sum_{u'<v'<w'\in\mathrm{CAS}}\left(\Delta\mathbf{F}_{u'v'w'}^\mathrm{corr}\right)_{ip}+\hdots \label{genfock:1} \\
    F_{up}^\mathrm{corr}&=&&\sum_{u'\in\mathrm{CAS}}\left(\mathbf{F}_{u'}^\mathrm{corr}\right)_{up}+\sum_{u'<v'\in\mathrm{CAS}}\left(\Delta\mathbf{F}_{u'v'}^\mathrm{corr}\right)_{up}+\sum_{u'<v'<w'\in\mathrm{CAS}}\left(\Delta\mathbf{F}_{u'v'w'}^\mathrm{corr}\right)_{up}+\hdots \label{genfock:2}
\end{align}
One key difference with respect to an RDM-based MBE is the fact that every increment will contribute to every element of the generalized Fock matrix. Explicit expressions for the individual blocks of the generalized Fock matrices that contribute to Eqs. \ref{genfock:1} and \ref{genfock:2} are provided in the SI.\\

Finally, we note how MBEs of tensors rather than scalars will add to the memory requirements. However, in the absence of any screening, closed-form $n$th-order approximations on par with those proposed in Ref. \citenum{Kaplan1995} may be used, which require saving only a single target property per MBE order. For instance, in the case of the generalized Fock matrix,
\begin{equation} \label{recursive}
    \mathbf{F}^{\mathrm{corr},(n)}=\mathbf{F}_\Sigma^{\mathrm{corr},(n)}-\sum_{m=1}^{n-1}\binom{M-n}{M-m}\mathbf{F}^{\mathrm{corr},(m)} \ ,
\end{equation}
we can construct this tensor from a sum of all current-order correlated Fock matrices, $\mathbf{F}_\Sigma^{\mathrm{corr},(n)}$, and its approximations at previous orders, $\mathbf{F}^{\mathrm{corr},(m)}$, with corresponding prefactors. In Eq. \ref{recursive}, $M$ denotes the total number of MOs in the expansion space.

\subsection{Orbital Optimization}

For the orbital optimization algorithm, we have adopted a mixed Super-CI and quasi-Newton approach. Super-CI is based on the generalized Brillouin theorem.\cite{Levy1968} Optimal parameters are obtained by solving a generalized eigenvalue problem defined in the basis of the reference wave function and all possible internally-contracted singly excited states. Exact solutions to the Super-CI problem are bound to be expensive, given how the evaluation of matrix elements in the Hamiltonian between singly excited states requires three-body RDMs over the active orbitals, and we have thus opted for a popular approximation that makes use of an effective one-electron Hamiltonian instead.\cite{roos1980_sci,malmqvist1990} Furthermore, to facilitate accelerated convergence, we have implemented a modified version of the direct inversion in the iterative subspace (DIIS) algorithm, as proposed by Kollmar and co-workers.\cite{kollmar2019_ptsci}\\

The Super-CI approach explicitly requires one-body RDMs. As touched upon in Sect. \ref{mbe_fci_casscf}, MBEs of RDMs may converge slowly, and early numerical experiments of our implementation showed how this can lead to stagnation of the Super-CI solver when approaching convergence. As an alternative, we have also implemented a quasi-Newton optimizer based on the limited-memory Broyden–Fletcher–Goldfarb–Shanno (L-BFGS) update.\cite{nocedal2006,fischer1992_qn} As suggested in Ref. \citenum{chaban1997_qn}, we use an approximate Hessian diagonal as preconditioner in order to avoid the computation of integrals that reference two nonactive indices. When considering active-active rotations, the exact Hessian diagonal of the active-active block is used; we note that our expression differs from the one presented in Ref. \citenum{levine2020casscf} but is equivalent to the one given in Ref. \citenum{Guo2021}, and the corresponding derivation is provided in the SI.\\

Since the BFGS step can easily lead an optimization too far out on the parameter hypersurface for the quadratic approximation to be valid, we perform an inexact line-search in compliance with Armijo's rule.\cite{armijo1966minimization} In particular, we require the energy to be sufficiently decreased with respect to the previous iteration. First, we try to accept the full BFGS step, and if it fails to fulfill Armijo's condition, we backtrack along the quasi-Newton direction until such criteria are met. The new back-tracked point is then provided as an argument of the minimum of a cubic model function, which is interpolated using the previously rejected points. The L-BFGS optimizer only relies on the MO gradient, and thus does not require the explicit evaluation of RDMs, but only a generalized Fock matrix, which, as discussed earlier, may exhibit a more robust convergence in an MBE. On the other hand, the convergence of the proposed quasi-Newton algorithm can be slow when starting far from the minimum, and we thus adopt a hybrid strategy, starting with a few Super-CI steps and then switching to the quasi-Newton code.\cite{malmqvist1990,toru2015_qn}\\ 

In general, regardless of whatever solver used, the bottleneck in the optimization of MOs will be the assembling of quantities required for the gradient evaluation, that is, both the inactive and active Fock matrices in Eqs. \ref{inact_fock_matrix} and \ref{act_fock_matrix} and the $\mathbf{Q}$ matrix in Eq. \ref{aux_matrix}. To remedy this, we exploit an optimized Cholesky decomposition of the involved two-electron integrals,\cite{Beebe1977,Koch2003} a technique which has already been employed in the context of CASSCF in previous work authored by some of us.\cite{nottoli2021second}

\section{Results and Discussion}

The CASCI step of all CASSCF calculations was performed using the open-source {\texttt{PyMBE}} code,\cite{pymbe} which is built upon the {\texttt{PySCF}} program.\cite{Sun2017, Sun2020} The Pipek-Mezey orbital localizations were also performed using {\texttt{PySCF}}. During this work, new features were added to {\texttt{PyMBE}} for the computation of one- and two-body RDMs and the generalized Fock matrix. The driver and CASSCF orbital optimization steps were performed using the {\texttt{CFOUR}} quantum chemistry suite of programs.\cite{cfour,matthews2020cfour} At each CASSCF iteration, an MBE is performed over the active orbitals in question to produce an approximation of either 1- and 2-RDMs or the generalized Fock matrix. Rather than canonical Hartree-Fock orbitals, we exploited unrestricted natural orbitals (UNOs) as a guess for the CASSCF procedure.\cite{pulay1988uhf,Bofill1989,Keller2015} In the course of the present study, we included all UNOs with occupation numbers between 0.01 and 1.99 in the active space.

\subsection{Calibration}\label{calib_sect}

To gauge the applicability of the MBE-CASSCF approach and to find optimal parameters that allow for a satisfactory compromise between accuracy and computational performance, we initially test the method on a set of three linear acenes in a standard cc-pVDZ basis set:\cite{Dunning1989} naphthalene, anthracene, and tetracene~\bibnote{A CCSD/cc-pVDZ geometry was used for the former system while the geometries for the latter two systems were taken from Ref. \citenum{Park2021}}. For all three systems, we correlate the $\pi$-orbitals resulting in (10,10), (14,14), and (18,18) active spaces, respectively. We follow a conventional notation of denoting the number of electrons and number of orbitals in said spaces by the first and second entries within the parentheses, respectively. For the two smallest systems, a conventional CASSCF calculation is still computationally viable while the latter tetracene system already demands an efficient large-scale implementation. The correlation energies and errors for all calculations in Section \ref{calib_sect} are provided in the SI.
In the following, we will discuss the following four main points: $(i)$ the effect of using truncated MBEs in the orbital optimization, and how the CASSCF energy can be improved with a final, accurate MBE-CASCI calculation; $(ii)$ the choice of MBE target, i.e., RDMs or the generalized Fock matrix, and how these affect the overall convergence, accuracy, and performance; $(iii)$ the importance of using localized active orbitals and the effect of choice of reference space on accuracy and convergence; and $(iv)$ the lack of exact invariance of the MBE-CASSCF energy with respect to active-active rotations, and whether to explicitly account for this or not.

\subsubsection{CASSCF MOs from Truncated MBEs}

\begin{figure}[htbp]
    \centering
    \includegraphics[width=0.8\textwidth]{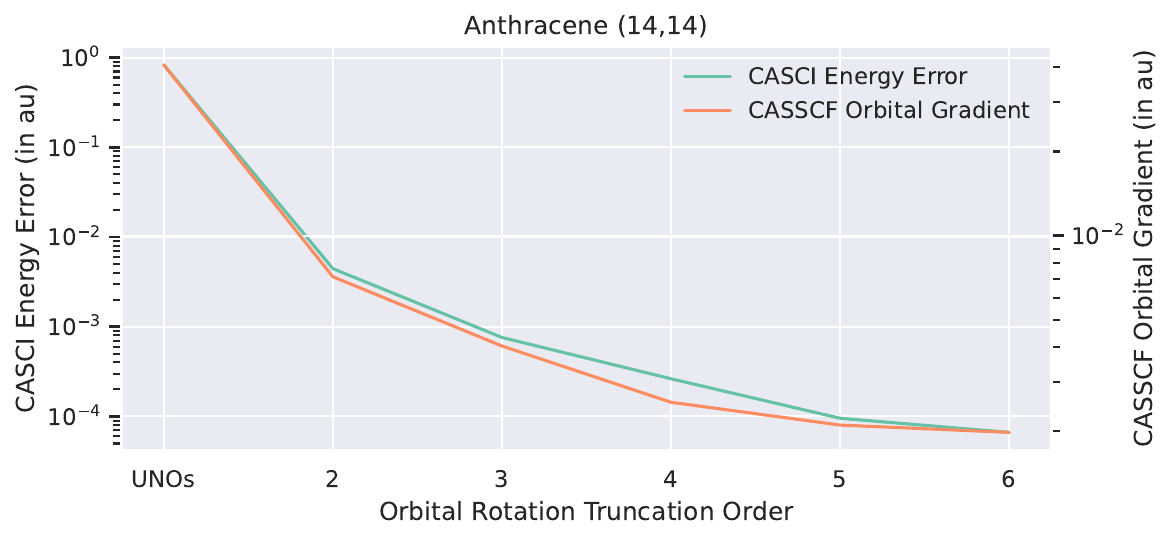}
    \caption{Error of an exact CASCI calculation and CASSCF orbital gradient of anthracene with a (14,14) active space for different orbitals in comparison to the exact CASSCF energy. The orbitals are either UNOs or orbitals from an MBE-CASSCF quasi-Newton orbital optimization where the MBE is truncated at different fixed orders.}
    \label{orb_rot_truncation}
\end{figure}
One idea suggested in earlier work on CASSCF algorithms that involve any sort of approximate CASCI solver is to differentiate between the quality of solutions used during the optimization of the CASSCF orbitals and in a final energy evaluation.\cite{Smith2017} Such an approach was tested for the anthracene molecule in Figure \ref{orb_rot_truncation} by computing the final energy using an exact CASCI solution on either simple UNOs or MOs from MBEs truncated at different fixed orders during the CASSCF orbital optimization (subject to no orbital screening).\\

From Figure \ref{orb_rot_truncation}, we note how the UNO starting guess can be significantly improved through MBE-CASSCF calculations truncated at increasing orders. As a result, the energy converges toward the exact CASSCF energy and the gradient correspondingly vanishes in the asymptotic limit. These results provide a convincing argument in favor of manually truncating MBEs at orders 4 or 5 during the orbital optimization, as the resulting orbitals will yield errors in total energies below chemical accuracy, on par with the observations made in Ref. \citenum{Smith2017}.\\

From a computational point of view, if MBEs are truncated without any screening protocol during the optimization of orbitals, we may evaluate a given target using Eq. \ref{recursive}, requiring only a single set of 1- and 2-RDMs or a single set of generalized Fock matrix blocks to be saved per MBE order. In all calculations to follow, MBEs will thus be truncated at order 5 during the orbital optimization. In general, final energy evaluations cannot be done in an exact manner (as in Figure \ref{orb_rot_truncation}) for larger active spaces, and will therefore be replaced by an MBE calculation subject to screening. We will further investigate the effect of such screening on the accuracy of the final MBE-CASCI energy in Sections \ref{target_section} and \ref{orbital_section}.

\subsubsection{Choice of MBE Target} \label{target_section}

\begin{figure}[htbp]
    \centering
    \includegraphics[width=\linewidth]{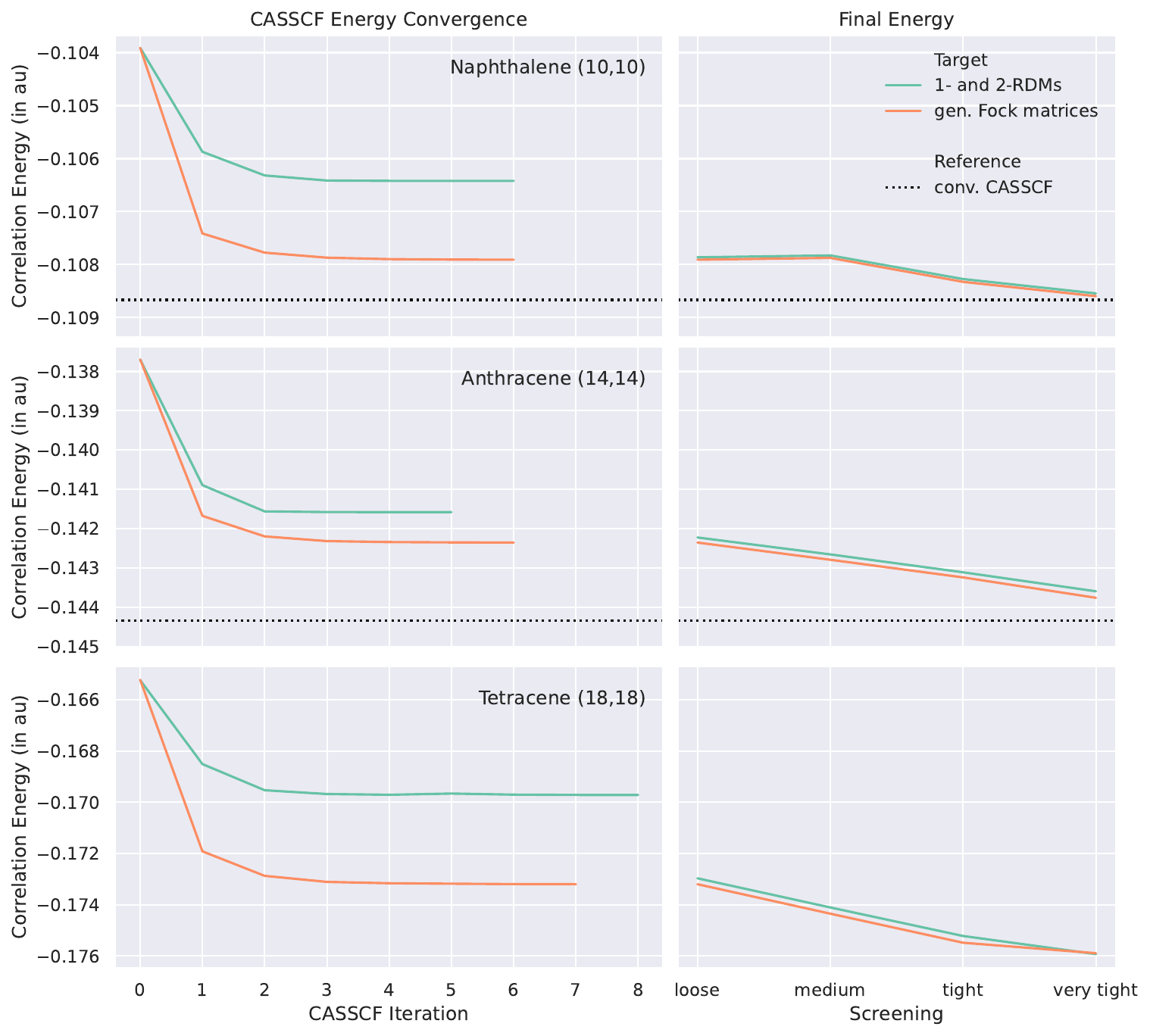}
    \caption{CASSCF energy convergence and final energy of naphthalene, anthracene and tetracene for expansions based on 1- and 2-RDMs and generalized Fock matrices. Pipek-Mezey localized orbitals are used as the initial guess for the active orbitals. The reference space involves those localized orbitals that have the greatest overlap with the UNOs with occupation numbers greater than 0.2 and smaller than 1.8.}\label{targets}
\end{figure}
As outlined in Sect. \ref{impl_sect}, we may target individual quantities by means of MBEs, namely, 1- and 2-RDMs and the generalized Fock matrix. In the leftmost panels of Figure \ref{targets}, we compare the convergence profiles of quasi-Newton orbital optimizations using either target. We note how the total energy can be readily evaluated from MBEs of 1- and 2-RDMs, while this is not possible from an expansion of the generalized Fock matrix. In the latter case, an expansion of the energy is instead done simultaneously at every iteration to keep track of energy convergence. In the left panel of Figure \ref{targets}, both optimizations start at the same energy at iteration 0 because the gradient and Hessian diagonal for the preconditioner is calculated from an expansion of RDMs in both implementations of MBE-CASSCF. While both orbital optimizations generally converge in a similar number of iterations, the implementation using the generalized Fock matrix converges to an energy that is much closer to the exact CASSCF result. 
This result is expected, as the MBE converges faster for the energy than for the RDMs. Furthermore, as the MO gradients are different between the two optimizations, a different convergence profile is not surprising.\\

\begin{table}
    \caption{Different screening parameters used in this work.}
    \label{screening}
    \begin{tabular}{ccc}
        \hline
        Screening & Start Order & Percentage \\
        \hline
        loose & 5 & 100\% \\
        medium & 5 & 30\% \\
        tight & 6 & 30\% \\
        very tight & 7 & 30\% \\
        \hline
    \end{tabular}
\end{table}
To appreciate the greater accuracy of the results obtained using an MBE of the Fock matrix, we can compare converged CASSCF results with the energies obtained in a final MBE-CASCI calculation based on the MBE-CASSCF orbitals. The rightmost panels of Figure \ref{targets} report final MBE-CASCI energies obtained using various screening parameters, cf. Table \ref{screening}. The screening scheme used throughout this work is orbital-based,\cite{Eriksen2021} and thus equivalent to the one used for the MBE-FCI calculation in Ref. \citenum{Eriksen2020a}. From a given order, the algorithm starts to remove a certain percentage of orbitals from the active space at every following order. The percentage of orbitals with the lowest maximum increment are removed from the expansion space. The expansion continues to increasingly higher orders, removing the least important orbitals from consideration at every order, until no valid increments can be produced from the remaining orbital space.\\

A comparison of the resulting final energies in the right panels of Figure \ref{targets} shows that a final MBE-CASCI calculation completely corrects for the error made by using approximate 1- and 2-RDMs in the orbital optimization, given how the orbitals obtained with either MBE target produce very similar final energies. From a computational point of view, MBEs of either target can be achieved using a closed-form, recursive implementation (Eq. \ref{recursive}), meaning that neither is memory limited, and either expansion produces very similar energies upon a final MBE-CASCI correction. However, the greater accuracy exhibited in the orbital optimization of MBEs of generalized Fock matrices motivates us to target this quantity in all quasi-Newton calculations to follow. Nevertheless, the accuracy of RDMs from MBEs is more than sufficient to perform a few initial super-CI steps and compute an initial approximated diagonal of the orbital rotation Hessian for the subsequent quasi-Newton optimization. 

\subsubsection{Orbital Localization and Choice of Reference Space} \label{orbital_section}

\begin{figure}[htbp]
    \centering
    \includegraphics[width=\linewidth]{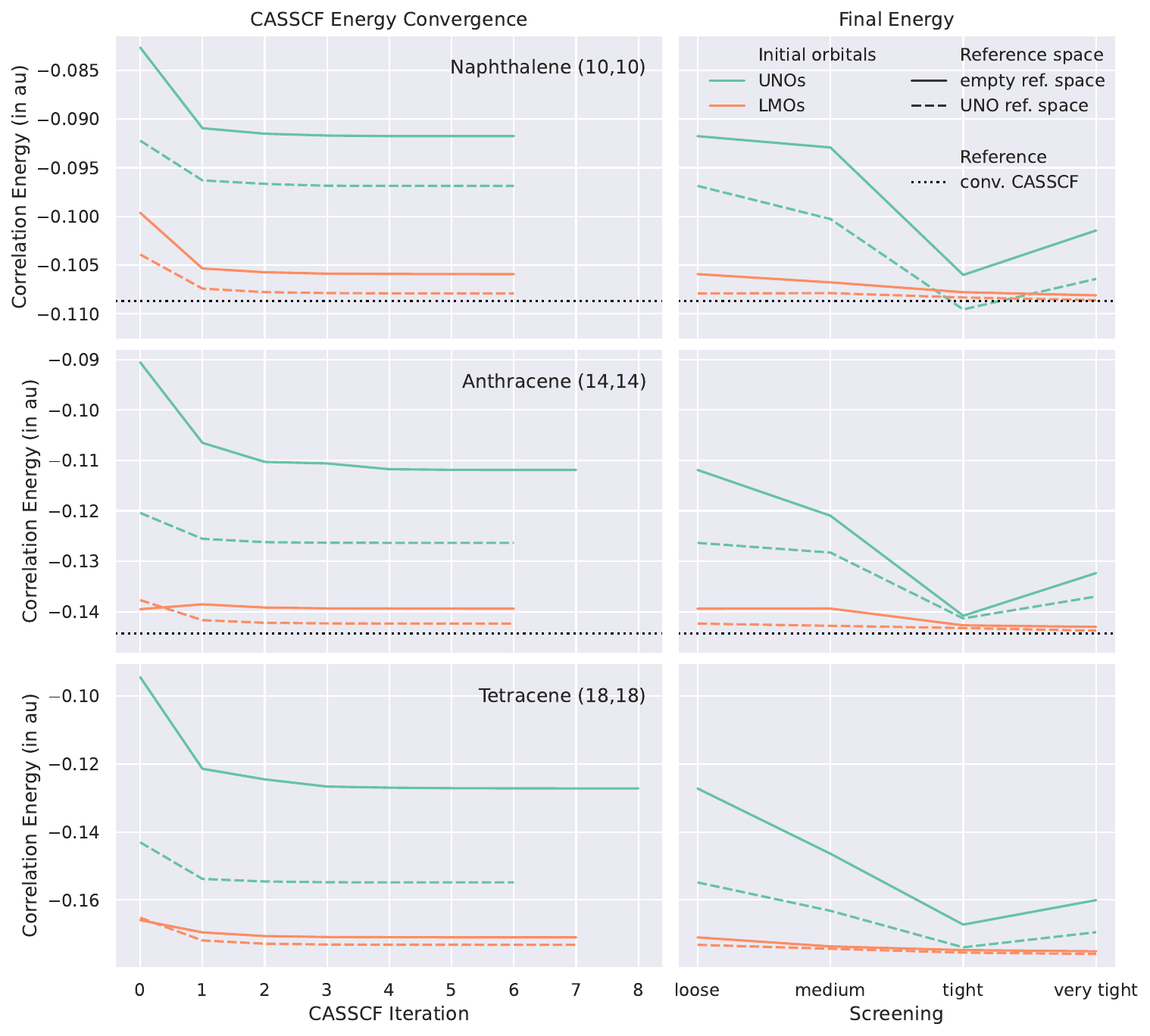}
    \caption{CASSCF energy convergence and final energy of naphthalene, anthracene, and tetracene for Pipek-Mezey localized orbitals (LMOs), unrestricted natural orbitals (UNOs), empty reference spaces, and references spaces based on UNO occupation numbers. Generalized Fock matrices are chosen as the MBE target in all these expansions.}\label{orbs_ref}
\end{figure}

Since MBE-CASSCF is not completely invariant under active-active MO rotations, the initial orbital guess can have a large effect on the accuracy of the obtained energies, and we next compare the convergence profiles of MBE-CASSCF optimizations and final MBE-CASCI energies using either UNOs or Pipek-Mezey localized MOs as starting points for the active space orbitals. The inactive and external orbital spaces are described by UNOs in both cases as this choice does not affect MBE-FCI convergence. We further investigate the effect of including a few orbitals in the reference space instead of using an empty reference space. The use of empty reference spaces provides a completely unbiased expansion, but is often plagued by a combinatorial explosion at higher orders necessary to converge many typical expansions. On the other hand, the choice of the reference orbitals requires a selection criterion, which introduces some arbitrariness into the overall procedure. Here, to select reference space orbitals, we use either all UNOs that have occupation numbers between 0.2 and 1.8, or, in the case of localized orbitals, the same number of localized MOs with the largest overlaps with this set of UNOs.
The results of these numerical experiments are reported in Figure \ref{orbs_ref}.\\

As has previously been observed for calculations of correlation energies, constructing MBEs from spatially localized orbitals tend to yield significantly more compact expansions and improved convergence onto exact results, e.g., in comparison to the delocalized natural orbitals used in this work. This is apparent by comparing the final MBE-CASSCF energies in the left panels of Figure \ref{orbs_ref} to the energy obtained in a final MBE-CASCI energy evaluation (rightmost panels of Fig. \ref{orbs_ref}) when using localized orbitals. The final MBE-CASSCF energies in the localized orbital basis are systematically lower and closer to the final MBE-CASCI energies calculated from these same orbitals by employing tighter screening criteria. The effects of localization are even greater on the final MBE-CASCI energies, where results, when based on UNOs, exhibit oscillations instead of rapid convergence upon tightening the screening. This behavior reflects the nonvariationality of the truncated MBE-FCI approach and localization of the initial active space orbitals will generally help to remedy such oscillations. We also note how an explicit choice of reference space, and therefore a focus of the expansion onto a specific set of strongly correlated orbitals, appears to be beneficial to the overall accuracy of the MBE-CASSCF method. Both orbital and energy convergences with respect to screening thresholds are significantly improved, which, in turn, facilitates chemical accuracy in energies obtained using tight screening parameters.

\subsubsection{Active-Active Orbitals Rotations}\label{act_act_rot_section}

As mentioned previously, MBE-CASSCF energies will not be exactly invariant under rotations of the active orbitals. To investigate the effects of the lack of such an invariance, we next report on MBE-CASSCF calculations where these have been explicitly included in the optimization procedure. Similar ideas have been proposed in the context of the iCASSCF method of Dang and Zimmerman\cite{Dang2021} and in other approximate CASSCF methods.\cite{Ma2017,levine2020casscf,Park2021,Park2021a,Yao2021,Smith2022} The resulting method is variational, in the sense that the energy is minimized with respect to all variational degrees of freedom, and the Hellmann-Feynman theorem can be applied, although the final energy is not an upper bound to the exact ground-state energy. However, the explicit consideration of active-active rotations is not entirely straightforward, given how it introduces certain redundancies in the parametrization of the wave function. Such redundancies can be eliminated to first order by removing singly excited determinants from the increment calculations during orbital optimization, effectively approximating a CASSCF solution without single excitations,\cite{Dang2021} before reintroducing the contribution of singles back into the final CASCI energy evaluation.\\

\begin{figure}[!htb]
    \centering
    \includegraphics[width=\linewidth]{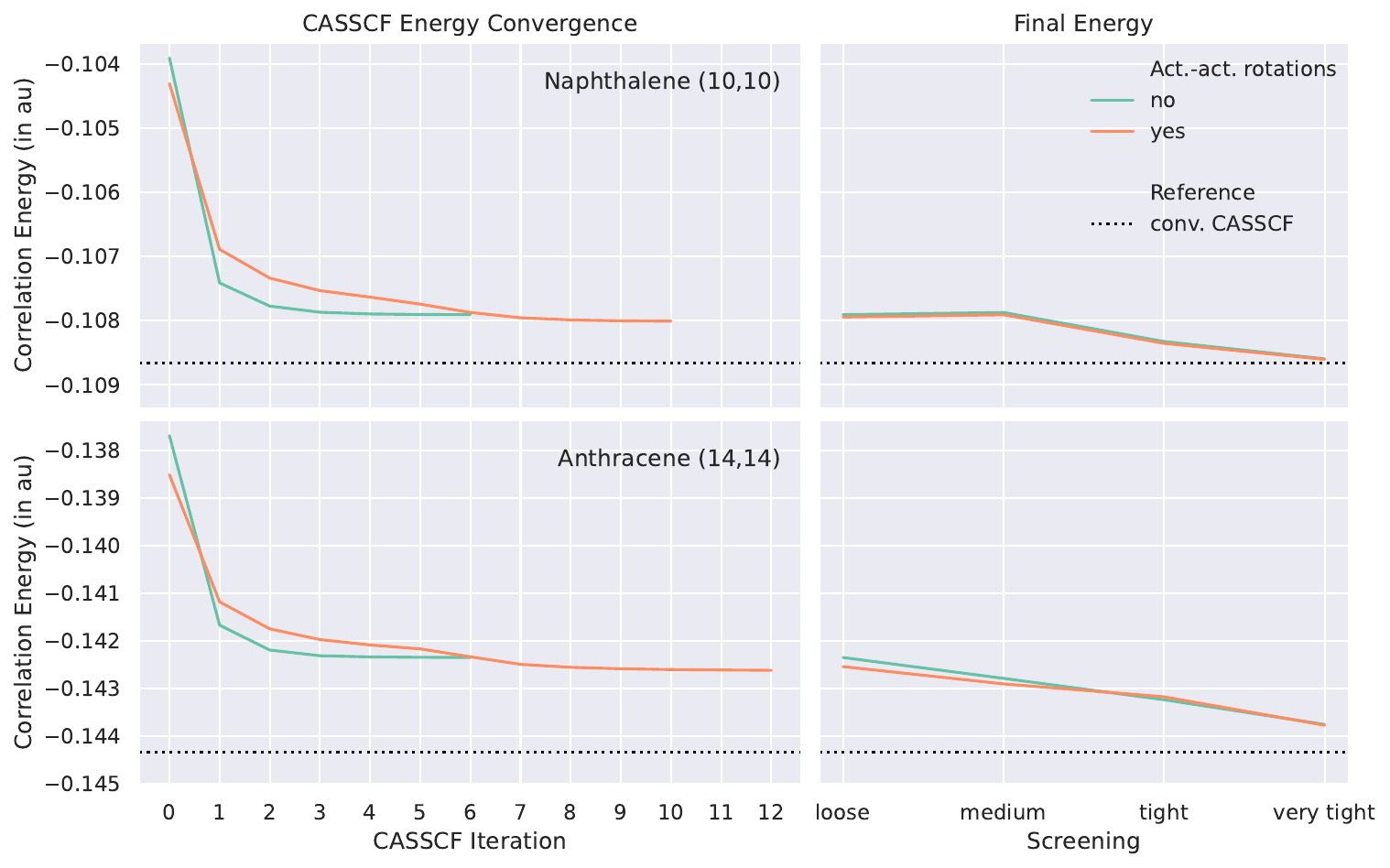}
    \caption{CASSCF energy convergence and final energy of naphthalene and anthracene with and without active-active rotations. Generalized Fock matrices are chosen as the target of these expansions, and Pipek-Mezey localized orbitals used as the initial guess for the active orbitals. The reference space contains those localized orbitals that have the greatest overlap with the set of UNOs that have occupation numbers greater than 0.2 and smaller than 1.8.}\label{act_act_rot}
\end{figure}
Figure \ref{act_act_rot} reports CASSCF optimization profiles and final MBE-CASCI energies for naphthalene and anthracene either with or without optimization of such rotations. We note how we were unable to converge the orbital optimization for the tetracene system when considering active orbital rotations, and we were furthermore unable to converge any calculations that did not make use of a localized orbital basis. The calculations that did indeed manage to converge did so to energies only marginally below those that neglect these rotations, albeit at the price of slower convergence. However, final MBE-CASCI energies are practically indistinguishable in both cases. Therefore, as in previous studies concerned with active-active rotations,\cite{levine2020casscf,Guo2021,Park2021,Park2021a,Smith2022} we conclude that the inclusion of these leads to unnecessary complications and increases to computational costs during the orbital optimization in MBE-CASSCF. Rather, high-quality orbitals are more appropriately optimized in a more cost-effective manner by performing MBE-CASSCF calculations in a suitable basis of localized active orbitals.

\subsection{Calculations on Fe(II) Porphyrin}

\begin{figure}[!htb]
    \centering
    \includegraphics[width=0.4\textwidth]{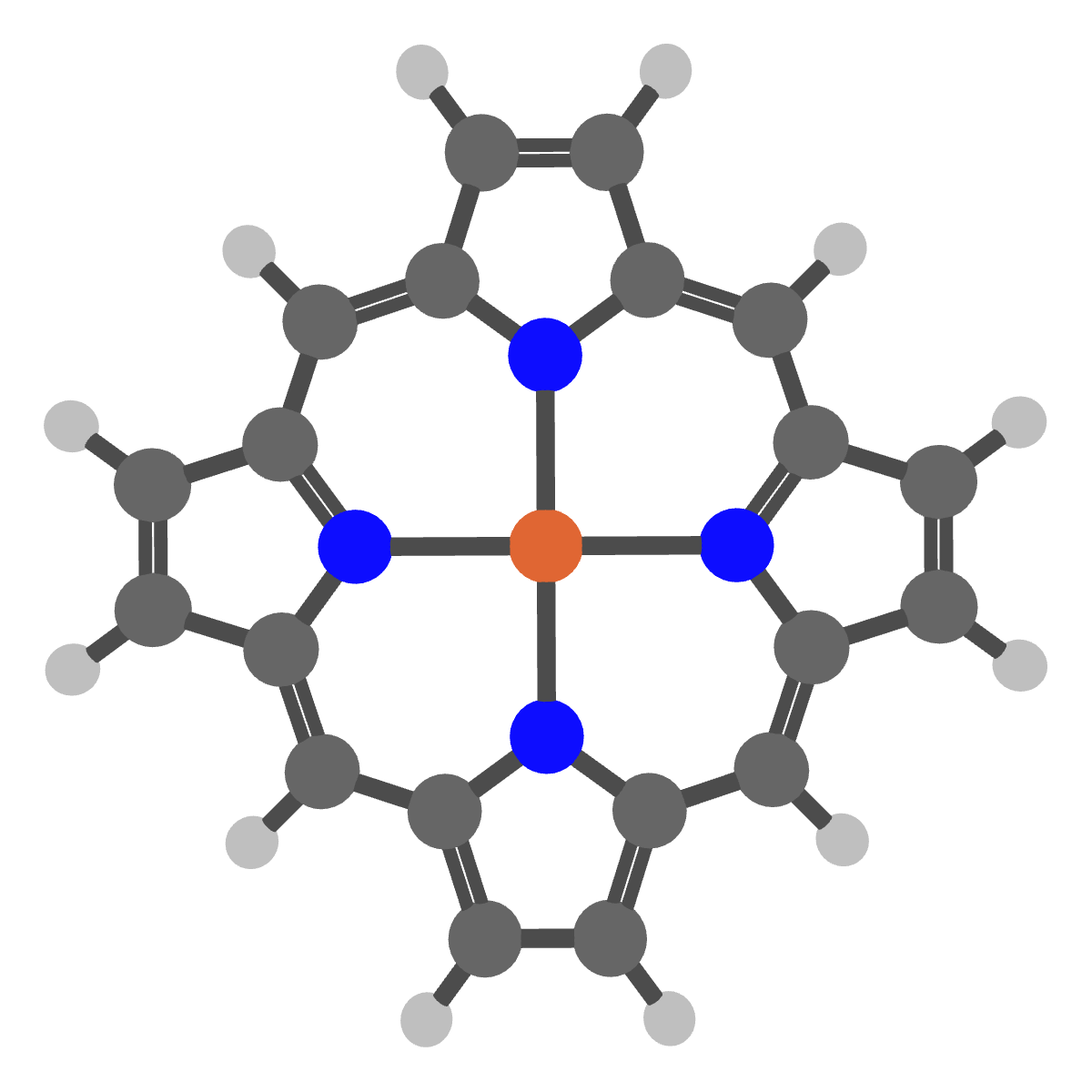}
    \caption{Molecular structure of Fe(II) porphyrin. Iron is depicted in orange, nitrogen in blue, carbon in dark gray, and hydrogen in light gray.}
    \label{fetpp}
\end{figure}

To test our MBE-CASSCF implementation and the parameters tuned via the numerical experiments described in Sect. \ref{calib_sect}, we next apply the method to a challenging system of great biological interest, namely, iron(II) tetraphenylporphyrin (Fe(II)tpp). Experimentally, the electronic structure of this iron porphyrin system has been characterized by means of M{\"o}ssbauer,\cite{collman1975synthesis, lang1978mossbauer} $^1$H-NMR~,\cite{goff1977nuclear, mispelter1980proton} and Raman spectroscopy,\cite{kitagawa1979resonance} and the temperature dependences of its magnetic susceptibility and paramagnetic anisotropy have also been investigated.\cite{boyd1979paramagnetic} All of these experimental studies agree on the fact that the electronic ground state of the system is of triplet spin, even if they disagree on the exact electronic structure. In a more recent study,\cite{tarrago2021experimental} based on a combination of leading experimental techniques, the ground-state configuration was found to be a mixture of two configurations of ${^3}A_{2g}$ and ${^3}E_g$ symmetry. On the other hand, theoretical studies on the unsubstituted Fe(II)tpp model system in Figure \ref{fetpp} have predicted different results based on the exact details of the given methodology used.\\

Early CASSCF/CASPT2 calculations predicted a quintet ground state,\cite{choe1998theoretical_a} while CCSD(T) calculations have predicted a triplet state.\cite{radon2014spin} The exact ordering of the electronic states in question is thus seemingly sensitive to both static and dynamic correlation, and, being a model system, other sources of discrepancies against experiment include missing phenyl groups, the pragmatic choice of the one-electron basis set employed in all of these studies, as well as possible environment and finite-temperature effects. Those reservations aside, the Fe(II) porphyrin system in Fig. \ref{fetpp} represents an interesting challenge to approximate CASSCF methods as the active space of MOs may be enlarged in a systematic manner. As such, the works by Olivares-Amaya and co-workers as well as Smith and co-workers have exploited a (44,44) active space, showing the ground state to be of triplet spin.\cite{olivares2015ab,Smith2017} Guo et al. have used a variety of chemically motivated active spaces; their calculations with larger active spaces correctly predict a triplet ground state.\cite{Guo2021} The leading hypothesis these days is that such large active spaces are able to accommodate a significant amount of dynamic correlation, thus leading to the correct ordering of the involved states.\\

To validate our methodology and offer possible new insights into the theoretical description of the electronic structure of this system, we perform calculations with a number of active spaces of increasing size. In contrast to previous work,\cite{olivares2015ab, Smith2017, levine2020casscf} the active spaces are again chosen systematically based on the UNOs; using the same criterion as in Sect. \ref{calib_sect}, that is, selecting all orbitals with occupation numbers between 0.01 and 1.99, yields an active space of 22 electrons in 22 orbitals. We next enlarge this initial (22,22) space to include the 30, 40, and 50 most correlated spatial orbitals according to the UNO occupations, under the restriction that an equal number of electrons is included in the reference space. A UNO criterion of occupation numbers between 0.2 and 1.8 gives rise to (6,6) and (8,8) reference spaces for the triplet and quintet states, respectively. Among these are the singly occupied orbitals and an equal number of occupied and virtual orbitals. The calculations were performed with a modest ANO-VDZP basis set using the same geometry as in Refs. \citenum{nagashima1986binitio} and \citenum{Manni2016}.\\

In all calculations, the MBE step used UNOs and localized orbitals as starting guess for the inactive/virtual and active orbitals, respectively. Due to the larger reference space for the quintet state, calculations beyond order 6 in the MBE became practically intractable in parallel for large active spaces, since the memory requirements of the individual CASCI calculations grew large. For this reason, the screening was restricted to consider only increments up until order 6 during the final CASCI energy evaluations. A comparison for the smallest (22,22) active space obtained using this type of screening and the three tightest of the fixed criteria of Table \ref{screening} is presented in Table S1 of the SI. As is obvious from those results, the manual termination of the MBE after order 6 will produce both total energies and a spin gap in near agreement with the results from the tighter two screening criteria, and these energies can be expected to be within chemical accuracy of the exact energy.\\

While the UNO criterion suggests a (6,6) reference space size for the triplet state, the accurate comparison of state energies arguably requires both states to be treated on as equal a footing as possible. For this reason, energies relying on both of these (6,6) and (8,8) reference spaces are compared in Table S2 of the SI, revealing energetic differences of less than $1 \  \text{m}E_{\text{h}}$. As such, and also due to computational efficacy, the UNO-derived (6,6) reference space will be used in the calculations of the triplet state for all studied active spaces.\\

\begin{figure}[!htb]
    \centering
    \includegraphics[width=0.9\textwidth]{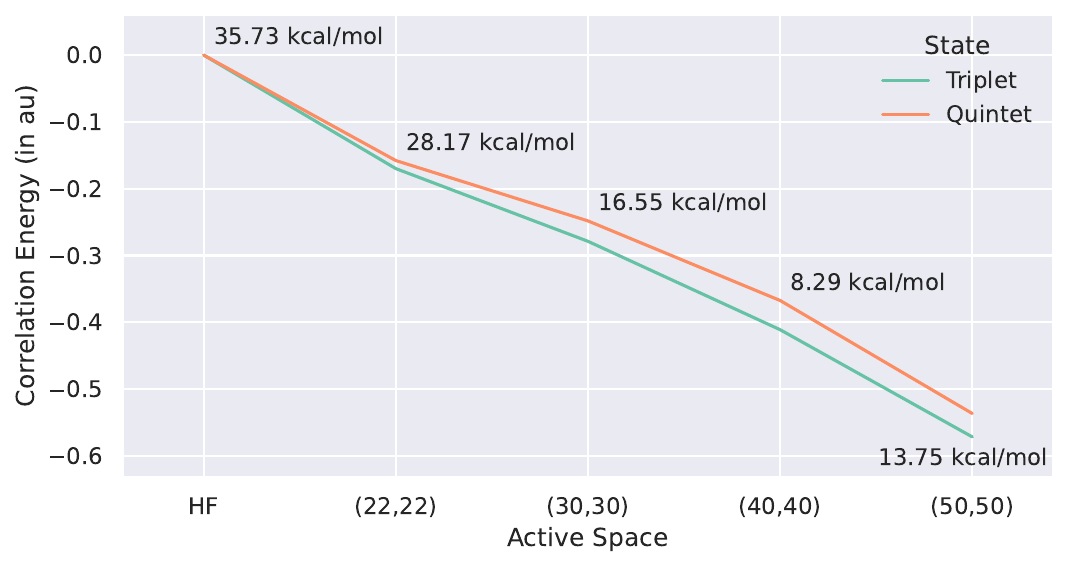}
    \caption{Correlation energies of the triplet and quintet states of the Fe(II) porphyrin system for different active spaces. The corresponding quintet-triplet spin gaps are annotated as text.}
    \label{fetpp_spin_gap}
\end{figure}
MBE-CASSCF correlation energies of both states (based on ROHF wave functions employing the coupling parameters suggested by Guest and Saunders~\cite{Guest1974}) and corresponding spin gaps for different active space sizes are reported in Figure \ref{fetpp_spin_gap}, with corresponding electronic energies reported in Table S3 of the SI. The correlation of the electrons within the orbitals of the active space leads to similar decreases of the electronic energies for both states upon an increase in the size of this space. As is obvious from the results in Fig. \ref{fetpp_spin_gap}, the energy of the triplet state generally changes more when correlating the electrons within a given active space, forcing the spin gap between the two states to close as more electrons and orbitals are successively added to the active space. As such, it is relatively easy to imagine a reversal of the state ordering in even larger active spaces. In line with the other theoretical studies on the Fe(II) porphyrin system,\cite{Manni2016,Smith2017,levine2020casscf} a possible reason that approximate CASSCF methods are unable to replicate the experimental ordering of the electronic states is due to the fact that limited active spaces cannot sufficiently account for the dynamic correlation present in the system. Phrased differently, the spin gap is so small that the significant approximations introduced by the native CASSCF method will not yield qualitatively correct results.\\

The MBE-FCI method is able to reliably converge the CASSCF optimization for large active spaces to an energy minimum, predicting an energy gap comparable with the ones reported in the literature.\cite{Smith2017,levine2020casscf} That being said, the energy gap for the (50,50) MBE-CASSCF calculation cannot be expected to be converged with respect to the size of the active space as the spin gap still fluctuates some upon adding more orbitals to the active space. While other approximations, such as, the missing phenyl rings or the limited one-electron basis set, could contribute to the qualitative disagreement with experiment, these fluctuations, also with respect to the actual choice of the active space orbitals themselves,\cite{olivares2015ab,Smith2017} can be explained by the missing correlation due to the remaining orbitals. In general, the neglected correlation should instead be treated more efficiently using dynamic correlation methods on top of an approximate CASSCF solution based on a more moderate active space. Within the MBE-CASSCF framework, this could potentially be achieved by replacing the final MBE-CASCI calculation with an MBE-FCI calculation, in which orbitals outside the active space are screened away relatively early on, so as to recover some of the dynamic correlation, while the static correlation within the active space is treated to the same degree. Such a strategy would drastically limit the additional cost of treating correlation outside the active space.

\section{Summary and Conclusions}

We have introduced the MBE-CASSCF method which couples two standard first-order CASSCF algorithms with a massively parallel implementation of the MBE-FCI method. The resulting algorithm is able to treat arbitrary and large active spaces by overcoming the steep scaling of the CASCI step in conventional CASSCF calculations. In the present study, we have used the MBE-CASSCF method to determine the energetic gap between the two lowest-lying electronic states of iron(II) porphyrin within active spaces comprising as many as 50 electrons in 50 orbitals.\\

The orbital optimization of our proposed algorithm can be efficiently accelerated and memory requirements significantly reduced through early termination of MBEs, while final energy evaluations will still produce results within chemical accuracy. We have shown how convergence onto conventional CASSCF may be improved through the use of compact orbital representations in the active space, such as, spatially localized orbitals, and we have discussed how the explicit internal optimization of active orbitals will generally lead to severe convergence issues for our algorithm and should thus be avoided. Such problems could potentially be remedied by an explicit coupling of CI coefficients and active-active orbital rotations.\cite{Kreplin2019, Yao2021} Another caveat of large-scale CASSCF calculations is the possible presence of multiple local minima, which has not been dealt with in the current implementation.\cite{levine2020casscf} In terms of possible avenues for future developments, the extension of our proposed MBE-based algorithm to molecular gradients would allow for geometry optimizations of statically correlated systems. In addition, final MBE-CASCI energy evaluations could be replaced by more complete MBE-FCI calculations, which would pose an intuitive alternative to multireference CI and perturbative methods for the treatment of larger active spaces.

\begin{acknowledgement}

JJE gratefully acknowledges two research grants, no. 37411 from VILLUM FONDEN (a part of THE VELUX FOUNDATIONS) and no. 10.46540/2064-00007B from the Independent Research Fund Denmark. IG and FL acknowledge financial support from ICSC-Centro Nazionale di Ricerca in High Performance Computing, Big Data, and Quantum Computing, funded by the European Union -- Next Generation EU -- PNRR, Missione 4 Componente 2 Investimento 1.4. The authors gratefully acknowledge the computing time granted on the Mogon II supercomputer by the Johannes Gutenberg-Universität Mainz (hpc.uni-mainz.de).

\end{acknowledgement}

\begin{suppinfo}

The supporting information (SI) presents additional details on 1- and 2-RDMs as well as generalized Fock matrices calculated from MBEs. Furthermore, analytical expressions for the generalized Fock matrix increments, the derivation for the active-active block of the orbital Hessian diagonal, and additional data for the calibration systems and the Fe(II) porphyrin system are provided.

\end{suppinfo}

\providecommand{\latin}[1]{#1}
\makeatletter
\providecommand{\doi}
  {\begingroup\let\do\@makeother\dospecials
  \catcode`\{=1 \catcode`\}=2 \doi@aux}
\providecommand{\doi@aux}[1]{\endgroup\texttt{#1}}
\makeatother
\providecommand*\mcitethebibliography{\thebibliography}
\csname @ifundefined\endcsname{endmcitethebibliography}
  {\let\endmcitethebibliography\endthebibliography}{}


\begin{mcitethebibliography}{89}
\bibitem[Not()]{Note-1}
A CCSD/cc-pVDZ geometry was used for the former system while the geometries for
  the latter two systems were taken from Ref. \citenum{Park2021}\relax
\mciteBstWouldAddEndPuncttrue
\mciteSetBstMidEndSepPunct{\mcitedefaultmidpunct}
{\mcitedefaultendpunct}{\mcitedefaultseppunct}\relax
\EndOfBibitem
\end{mcitethebibliography}
\end{document}